\documentclass[aps,prd,superscriptaddress,long,notitlepage,balancelastpage,nofootinbib,floatfix,twocolumn]{revtex4-2}
\pdfoutput=1
\usepackage{amsmath,mathtools,amssymb,amsthm,amsxtra,overpic,bbm,epsfig,subfigure,url,bm}
\usepackage{hyperref}
\usepackage{mathrsfs}
\usepackage{color,xcolor}
\usepackage{comment}
\usepackage{float}
\usepackage{enumitem}
\usepackage{slashed}
\usepackage{multirow}
\usepackage{appendix}

\usepackage[normalem]{ulem}

\DeclareUnicodeCharacter{03BC}{\ensuremath{\mu}}
\DeclareUnicodeCharacter{2032}{\ensuremath{'}}
\DeclareUnicodeCharacter{2009}{\,}
\DeclareUnicodeCharacter{3B1}{\ensuremath{\alpha}}
\definecolor{nicered}{rgb}{0.5,0.,0.}
\definecolor{nicegreen}{rgb}{0.,0.5,0.}
\definecolor{niceblue}{rgb}{0.,0.,0.5}
\hypersetup{
	colorlinks=true,
	linkcolor=black,
	filecolor=nicegreen,      
	urlcolor=niceblue,
	citecolor=nicered,
}

\makeatletter
\newcommand*{\balancecolsandclearpage}{%
	\close@column@grid
	\cleardoublepage
	\twocolumngrid
}
\makeatother
\begin{document}
\title{{\it \textbf{Flavor Matters, but Matter Flavors}}: \\
Matter Effects on Flavor Composition of Astrophysical Neutrinos}

\author{\bf P. S. Bhupal Dev}
\email[E-mail:]{bdev@wustl.edu}
\affiliation{Department of Physics and McDonnell Center for the Space Sciences, Washington University, St.~Louis, Missouri 63130, USA}

\author{\bf Sudip Jana}
\email[E-mail:]{sudip.jana@mpi-hd.mpg.de}
\affiliation{Harish-Chandra Research Institute, A CI of Homi Bhabha National Institute, Chhatnag Road, Jhunsi, Prayagraj 211 019, India}
\affiliation{Max-Planck-Institut f{\"u}r Kernphysik, Saupfercheckweg 1, 69117 Heidelberg, Germany}

\author{\bf Yago Porto}
\email[E-mail:]{yago.porto@ufabc.edu.br}
\affiliation{Centro de Ciências Naturais e Humanas, Universidade Federal do ABC, 09210-170, Santo André, SP, Brazil}
\affiliation{Instituto de F{\'i}sica Gleb Wataghin, Universidade Estadual de Campinas, 13083-859, Campinas, SP, Brazil}

\begin{abstract}
We show that high-energy astrophysical neutrinos produced in the cores of heavily obscured active galactic nuclei (AGNs) can undergo strong matter effects, thus significantly influencing their source flavor-ratios. In particular, matter effects can completely modify the standard interpretation of the flavor ratio measurements in terms of the physical processes occurring in the sources (e.g., $pp$ versus $p\gamma$, full pion-decay chain versus muon-damped pion-decay). We contrast our results with the existing flavor ratio measurements at IceCube, as well as with projections for next-generation neutrino telescopes like IceCube-Gen2. Signatures of these matter effects in neutrino flavor composition would not only bring more evidence for neutrino production in central AGN regions, but would also be a powerful probe of heavily Compton-thick AGNs, which escape conventional observation in $X$-rays and other electromagnetic wavelengths.
\noindent 
\end{abstract}
\maketitle
\section{Introduction}
The discovery of high-energy neutrinos (HENs) in the TeV--PeV range by the IceCube Neutrino Observatory~\cite{IceCube:2013cdw, IceCube:2013low} has commenced a new era in Neutrino Astrophysics~\cite{Ahlers:2018fkn}. Follow-up observations by IceCube~\cite{IceCube:2014stg, IceCube:2015gsk, IceCube:2015qii, IceCube:2016umi, IceCube:2020wum, IceCube:2020fpi, IceCube:2021uhz}, as well as by ANTARES~\cite{ANTARES:2017srd} and Baikal-GVD~\cite{Baikal-GVD:2022fis}, have detected several  diffuse HEN events. The origins of these astrophysical neutrinos remain largely unknown~\cite{Kurahashi:2022utm, Troitsky:2023nli}. There is no strong anisotropy in the observed flux, so it is likely dominated by extragalactic sources, with only a subdominant galactic-component~\cite{Ahlers:2015moa, IceCube:2023ame}. Despite extensive multi-messenger observational campaigns involving neutrinos, cosmic-rays, gamma-rays, and gravitational waves~\cite{Greus:2021gba, Guepin:2022qpl}, the astrophysical sources of most of the extragalactic neutrino events still remain unaccounted for. 

Nevertheless, there is tantalizing evidence of a handful of extragalactic point-sources~\cite{IceCube:2018cha, IceCube:2018dnn, Rodrigues:2020fbu, IceCube:2022der, ANTARES:2023lck} -- the active galaxies TXS 0506+056, NGC 1068 and PKS 1424+240 being the most significant, albeit limited to TeV energies and  track events only. The fact that {\it all} these identified sources are  active galactic nuclei (AGNs) strongly indicates that AGNs are among the most promising candidate sources~\cite{IceCube:2021pgw, Murase:2022feu}. Among the possible AGN types, $\gamma$-ray blazars~\cite{Antonucci:1993sg, Urry:1995mg}, which are characterized by jets directed nearly towards the Earth, are limited to deliver only $\lesssim 20\%$ of the flux~\cite{IceCube:2016qvd}. In this context, hidden or obscured AGNs, with much weaker jets and suppressed $\gamma$-ray emission, are gaining increasing attention as plausible HEN factories~\cite{Murase:2015xka, Murase:2019vdl, IceCube:2021pgw, Inoue:2022yak, Fang:2022trf, Halzen:2023usr}. There is some evidence to back this up from the recent multiwavelength discovery of a large population of obscured AGNs~\cite{Hickox:2018xjf, Fermi-LAT:2018lqt, Li:2019zqc, 2020ApJ...897..160L, Carroll:2023mos, Yan:2023kuc, Signorini:2023egg, Lyu:2023ojm}. 

\begin{figure*}[t!]
\includegraphics[width=0.89\textwidth]{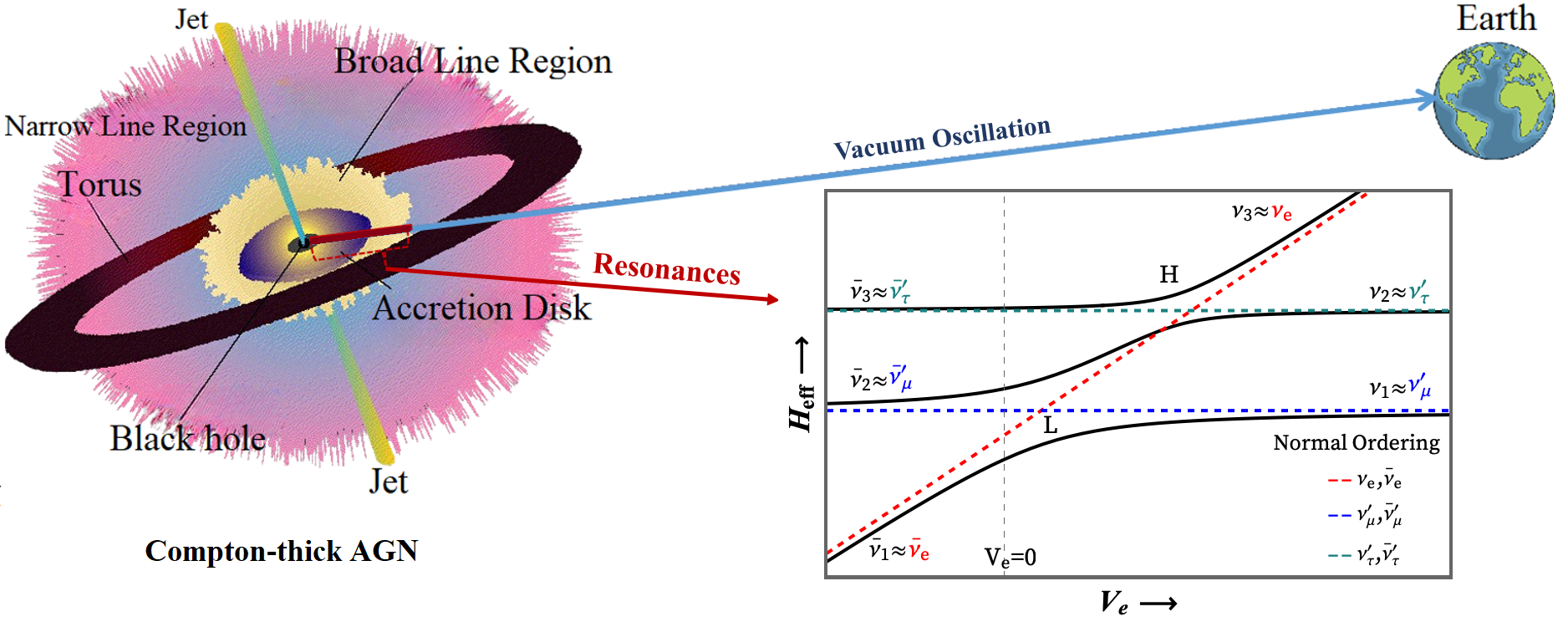}
  \caption{Schematic illustration of neutrino propagation in AGNs. Neutrinos are produced close to the central BH, where densities are very high and ME is important. As a consequence, neutrinos undergo MSW conversions ($H$ and $L$-resonances) before they reach the vacuum and propagate to the Earth. The energy levels of a neutrino system propagating in a slowly decreasing density profile are also shown ({see Appendix~\ref{app:energy})}.}
\label{Simplified_Fig}
\end{figure*}

In this scenario, neutrinos and $\gamma$-rays should be produced in very compact and dense regions, close ($\ll 1$ pc)  to the central supermassive black hole (BH), from where neutrinos escape but $\gamma$'s do not; see Fig.~\ref{Simplified_Fig}. The evidence for NGC 1068 as a neutrino source~\cite{IceCube:2022der} and subsequent studies modeling its emission~\cite{Murase:2022dog} further strengthens the case for obscured AGNs as the main astrophysical neutrino sources. 

In this paper, we show for the first time that if neutrinos are indeed produced in the central AGN regions, they might cross a dense medium with column-density large enough for strong matter effects (ME) to kick in~\cite{Lunardini:2000swa} while leaving the AGN core. 
As a consequence, the flavor content of the astrophysical neutrino-flux would be modified as compared to the usual expectation~\cite{Learned:1994wg} that takes only vacuum oscillations (VO) into account. This result has far-reaching implications for the correct interpretation of the flavor-ratio measurements at current~\cite{IceCube:2015gsk, IceCube:2023fgt} and future~\cite{IceCube-Gen2:2023rds} neutrino telescopes -- one of the essential tools to understand the HENs. 

ME for astrophysical neutrinos have been previously discussed in the context of chocked jets in supernovae (SNe) and gamma-ray bursts (GRBs)~\cite{Mena:2006eq, Razzaque:2009kq, Sahu:2010ap, Varela:2014mma, Xiao:2015gea, Carpio:2020app, Xu:2022wzh}. However, no correlations between HEN events and SNe or GRBs were found in dedicated searches, disfavoring these scenarios~\cite{IceCube:2017amx, ANTARES:2020vzs, IceCube:2022rlk, IceCube:2023esf}. Here, for the first time, we analyze the ME on neutrino flavor-conversion in the AGN environment, which, as discussed above, remains the most promising source type. 

\vspace{0.05in}

\section{Neutrino production in AGNs} AGNs are natural candidates for particle accelerators where neutrinos are created by hadronuclear interactions of accelerated protons with ambient protons ($pp$) or by photohadronic interactions of protons with ambient photons ($p \gamma$)~\cite{Murase:2015xka, Murase:2019vdl, Inoue:2022yak, Fang:2022trf, Halzen:2023usr}. 
Both $pp$ and $p \gamma$ interactions produce high-energy pions that subsequently decay to neutrinos, $\pi^{\pm} \rightarrow \mu^{\pm} + \nu_\mu/\Bar{\nu}_\mu \rightarrow e^{\pm} + \Bar{\nu_e}/\nu_e + \nu_\mu +\Bar{\nu}_\mu$, with source flavor-ratio $(f_{\nu_e+\Bar{\nu}_e},f_{\nu_\mu+\Bar{\nu}_\mu},f_{\nu_\tau+\Bar{\nu}_\tau})_{S}=(1:2:0)_S$\footnote{Antineutrinos are indistinguishable from neutrinos in the IceCube detector, so they are summed up in computing flavor-ratios. The only exception is the Glashow Resonance~\cite{Glashow:1960zz, Bhattacharya:2011qu, Huang:2023yqz, Liu:2023lxz}, but with poor statistics (only one candidate event so far~\cite{IceCube:2021rpz}).} which then propagate vast cosmic distances to reach the Earth. It is usually assumed that due to VO, neutrino-fluxes at Earth are shared approximately equally among the three flavors: $(f_{\nu_e+\Bar{\nu}_e},f_{\nu_\mu+\Bar{\nu}_\mu},f_{\nu_\tau+\Bar{\nu}_\tau})_{\oplus} =(1:1:1)_{\oplus}$~\cite{Learned:1994wg}, as predicted by the observed neutrino mixing angles~\cite{deSalas:2020pgw, Esteban:2020cvm, Capozzi:2021fjo}. Large deviations from the total flavor equipartition are possible. For instance, if the muon from $\pi$-decay rapidly loses energy due to environmental interactions (e.g. synchrotron emission) before decaying (muon-damped case)~\cite{Rachen:1998fd, Kashti:2005qa, Kachelriess:2007tr, Hummer:2010ai, Winter:2014pya}, the produced flavor-ratio is $(0:1:0)_S$ which translates to $(4:12:9)_{\oplus}$. Or, if the source somehow injects a nearly pure neutron flux~\cite{Anchordoqui:2003vc, Anchordoqui:2014pca}, the source flavor-ratio is $(1:0:0)_S$, which translates to $(14:4:7)_\oplus$.

\vspace{0.05in}

\section{Influence of AGN matter}
The main point of this paper is that neutrino flavor-conversion {\it is} necessarily affected by the high matter densities encountered in the AGN cores where neutrinos are produced.  We specifically focus on obscured AGNs~\cite{Hickox:2018xjf}, where neutrinos are produced in the vicinity of the central BH, at radial distances of 10--100 Schwarzschild radii ($R_s$)~\cite{Inoue:2019yfs, Murase:2022dog, Halzen:2023usr} and go through regions filled with relatively dense gas such as the upper layers of the BH accretion disk (AD), the torus, and the Broad Line Region (BLR) formed by gas that is bounded by the gravitational influence of the central BH, and emits optical/ultraviolet broad emission lines~\cite{2006LNP...693...77P, 2009NewAR..53..140G, 2018Natur.563..657G, 2017A&A...607A..32B, 2021MNRAS.502.3855L},   see Fig.~\ref{Simplified_Fig}. To give a sense of the scales involved here, the host galaxy is roughly 1 kpc across, the inner (outer) radius of the torus is about 1 (10) pc from the center, and neutrinos are produced close to the AD, $\sim 10^{-5}$ pc from the central region, equivalent to $10R_s$ for a BH of mass $10^7 M_\odot$~\cite{Murase:2022dog, Woo:2002un, Panessa:2006sg}. In the AD and inner parts of BLR, particle number densities are inferred to reach up to $10^{19} \, {\rm cm}^{-3}$ or even higher, given the limited observational data and the current theoretical understanding of the AGNs~\cite{Garcia:2018ckp, Jiang:2019xqn, Jiang:2019ztr}. In the outer parts of the BLR, where it meets the torus, densities drop to $10^{9} - 10^{11}\, \text{cm}^{-3}$~\cite{2016ApJ...831...68A,2017FrASS...4...19A, 2018ApJ...856...78A}. As demonstrated below,  neutrinos can undergo Mikheyev-Smirnov-Wolfenstein (MSW) resonant flavor-conversion~\cite{Wolfenstein:1977ue, Mikheyev:1985zog, Mikheev:1986wj, Mikheev:1987jp} on their way out from the AGN as the matter density decreases from the inner to the outer BLR region and the resonance criteria are met.

\vspace{0.05in}
\section{BLR geometry and density profile} 
{The BLR structure and geometry is not fully understood.}  
The BLR geometry is assumed to be either a flared disk-shaped or spherical~\cite{2006LNP...693...77P, 2009NewAR..53..140G, 2018Natur.563..657G, 2017A&A...607A..32B, 2021MNRAS.502.3855L}. It extends from the upper AD up to the inner edge of the torus. {The BLR structure, on the other hand, is usually depicted in the literature as either made out of discrete clouds or a continuous gas. A number of earlier works have discussed the issue of continuous versus discrete BLR (see Ref.~\cite{Laor:2005gg} and  references therein). The first notion of the structure of the BLR came from an analogy with galactic systems of clouds, especially the Crab Nebula system. This is somewhat expected since the first astrophysicists to examine the physics of the BLR came from the field of nebular physics in galactic sources. In addition, early estimations using the physics of discrete clouds were compatible with observations of the column density of X-ray absorbers in AGNs. Therefore, it was natural to assign this absorption to discrete BLR clouds~\cite{2006LNP...693...77P}. For this reason, many theoretical models were proposed to explain the origin and configuration of the clouds, including tidally disrupted stars~\cite{1992ApJ...385..108R}, star-disk collisions~\cite{1994ApJ...434...46Z}, and gravitational instability in the outer AD~\cite{Collin:2001ni}. 
Nevertheless, the physical viability of these ideas remains to be proven.}

{Among the several models to explain the formation and of the BLR clouds, one considered particularly promising was the model of bloated stars, where clouds are formed by supergiant stars that are further bloated by the ionizing radiation field. This model gives a natural explanation for how the cloud structure is maintained~\cite{1980MNRAS.190..757E}. Observational evidence from the least luminous known type I AGN (NGC4395), however, rules out the bloated star possibility, and at the same time reinforced the possible smooth and continuous nature of the BLR~\cite{Laor:2005gg}. The argument is the following: The size of the BLR is proportional to a power of the optical luminosity, $R_{\rm BLR} \propto L_{\rm opt}^{0.6}$. Therefore, NGC4395, being the least luminous, has a relatively small BLR size of $R_{\rm BLR} \sim 10^{14}$ cm, which turns out to be of the order of the predicted size of bloated stars~\cite{Alexander:1994vw}. As a result, one would naturally expect only a handful of bloated stars to compose the BLR of NGC4395. However, with such a small number of clouds, the BLR should exhibit significant amplitude fluctuations in the emission-line profiles. This is because the observed broad lines are the combined emission from all clouds, with each individual cloud contributing a narrow line. Statistical fluctuations in the cloud distribution per velocity bin would introduce irregularities in the emission-line profile, which scale as $N_c^{-1/2}$ (assuming a random distribution of cloud velocities), where $N_c$ represents the number of clouds. Consequently, an upper limit on the size of these fluctuations can be interpreted as a lower limit on $N_c$. In the case of NGC 4395, it has been shown that $N_c > 10^4$~\cite{Laor:2007hg}. This result contradicts the expectation of only a handful of clouds, as predicted by the bloated star scenario. 
Alternative scenarios, in which clouds are smaller than the bloated star size, can also be ruled out by other arguments~\cite{Laor:2005gg}. Following these arguments, we assume the BLR to be a smooth and continuous gas~\cite{2017A&A...607A..32B}, composed mostly of neutral Hydrogen, although a small deviation from this assumption is possible due to the partial ionization~\cite{2009NewAR..53..140G} without changing significantly our conclusions. Even if we take discrete BLR clouds, under certain conditions, our main results are still valid; see Appendix~\ref{app:BLR}.}

Assuming neutrinos are produced in outer AD\footnote{If neutrinos are produced outside the AD, matter effects may still be relevant, especially in the presence of a geometrically and optically thick BLR.} and then travel through the BLR, we can describe the electron number density profile encountered by neutrinos using a power law, as originally proposed in Ref.~\cite{1993ApJ...404L..51N} and broadly supported by detailed simulations and comparison with observational data~\cite{, 2016ApJ...831...68A, 2017FrASS...4...19A, 2018ApJ...856...78A}:
\begin{equation} \label{electron-density}
    n_e (r)= (10^{9} \, \text{cm}^{-3}) \left( \frac{0.1~\text{pc}}{r} \right)^\beta ,
\end{equation}
normalized to have $n_e=10^{9}\, \text{cm}^{-3}$ at the outer edge of the BLR ($r=0.1$ pc) and $n_e=10^{21}\, \text{cm}^{-3}$ in upper AD ($r = 10^{-5}$ pc) for $\beta=3$. We assume that Eq.~\eqref{electron-density} is valid for $r \gtrsim 10^{-5}$ pc ($r \gtrsim 10 R_s$ for a BH mass $10^{7} M_\odot$, such as in NGC 1068~\cite{Murase:2022dog}) so that the profile describes the whole range of densities from where neutrinos are produced to the end of the BLR. 

As for the power-law density slope $\beta$ in Eq.~\eqref{electron-density},  Refs.~\cite{2017FrASS...4...19A, 2018ApJ...856...78A} presented line luminosity radial profiles for a range of $\beta=0.5-3$ to simulate the photoionization process in the BLR gas and compared against the observed spectral shapes from multi-wavelength campaigns.  The density normalizations lower than the values chosen there do not reproduce the observed intermediate line emission. Similarly, the power law density distributions chosen there yield continuous line emissivity profiles with prominent intermediate line emission
component in permitted lines H$\beta$, He II and Mg II, independent of the density slopes and the spectral
radiation shapes adopted. Even though the power law density distribution of clouds may not fully reflect realistic situation in AGNs, it is sufficient for the purpose of our work. For concreteness, we choose to work with a conservative choice of $\beta=3$ and comment later on the consequences of modifying this assumption on our result. 

\vspace{0.05in}

\section{Evolution Hamiltonian} Neutrino flavor evolution can be described by the Schr\"{o}dinger  equation 
\begin{equation}
    i \frac{d}{d r} \nu=\mathcal{H}_{\rm eff} \nu \, ,
\end{equation}
where $\nu=(\nu_e,\nu_\mu,\nu_\tau)^T$ is the flavor state, $r$ is the radial coordinate in a system that has the BH in the center with neutrinos traveling outward through the BLR (Fig.~\ref{Simplified_Fig}) and $\mathcal{H}_{\rm eff}$ is the effective flavor Hamiltonian in presence of matter: 
\begin{equation} \label{H}
    \mathcal{H}_{\rm eff}=\frac{1}{2 E} U\left(\begin{array}{ccc}0 & 0 & 0 \\ 0 & \Delta m_{21}^2 & 0 \\ 0 & 0 & \Delta m_{31}^2\end{array}\right) U^{\dagger}+V_e\left(\begin{array}{ccc}1 & 0 & 0 \\ 0 & 0 & 0 \\ 0 & 0 & 0\end{array}\right).
\end{equation}
The first term on the right-hand side governs VO. The Pontecorvo–Maki–Nakagawa–Sakata (PMNS) matrix $U$ is parametrized in terms of three mixing angles $\theta_{ij}$ and a Dirac CP phase, $\Delta m_{j1}^2$'s are the mass-squared differences, and $E$ is the energy~\cite{ParticleDataGroup:2022pth}. The second term contains the matter potential $V_e(r)=\sqrt{2} G_F n_e(r)$, where $G_F$ is the Fermi constant, that affects only electron neutrinos via charged-current weak interaction $\nu_e + e^- \rightarrow \nu_e + e^-$~\cite{Wolfenstein:1977ue}. Neutral-current interactions give the same contributions to all flavors and do not impact the flavor-conversion. For antineutrinos, the sign of the matter potential flips: $V_e \rightarrow -V_e$~\cite{Linder:2005fc}. 

\vspace{0.05in}

\section{Flavor conversion} Neutrino flavor-conversion in a varying density profile such as Eq.~\eqref{electron-density} can happen at the high ($H$) and low ($L$) resonant layers~\cite{Dighe:1999bi}. The average number-density, $n_e^{\rm res}$, corresponding to the resonant layers is given by
\begin{equation} \label{resonance-conditions}
 \sqrt{2} G_F n_e^{\rm res}=  \frac{\Delta m^2_{i1} }{ 2 E} \cos 2 \theta_{1i} ,   
\end{equation}
where $L$ ($H$) corresponds to $i=2$ (3). Eq.~\eqref{resonance-conditions} is derived in the two-flavor approximation but is valid for the three-flavor system in Eq.~\eqref{H}, provided the resonant layers factorize and act independently~\cite{Dighe:1999bi}. Numerically, $n_e^H \approx 10^{20}\text{cm}^{-3} (100\text{ TeV}/E)$  and $n_e^L \approx 10^{18}\text{cm}^{-3} (100\text{ TeV}/E)$, with the oscillation parameters taken from Ref.~\cite{deSalas:2020pgw}. In addition, we can estimate the width of the resonant layer as $\Delta n_e^{\rm res} \approx n_e^{\rm res} \tan 2 \theta_{1i}$~\cite{Mikheev:1987jp}. It is necessary that neutrinos cross the entire resonant layer adiabatically for efficient conversion. The level of adiabaticity of the resonant layer is measured by
\begin{equation} \label{adiabaticity}
    \gamma = \frac{\Delta m^2_{i1} \sin^2 2 \theta_{1i}}{2E \cos 2 \theta_{1i}} \left( \frac{1}{n_e^{\rm res}} \left| \frac{d n_e}{d r} \right|_{\rm res} \right)^{-1},
\end{equation}
and $\gamma > 1$ means adiabatic propagation~\cite{Dighe:1999bi}.

If we assume the profile in Eq.~\eqref{electron-density} and neutrinos are produced at $10 R_s \sim 10^{-5}$ pc, then both resonance densities are met during propagation for $E>10$ TeV as $n_e(10^{-5}\ \text{pc})\simeq 10^{21}\text{cm}^{-3} > n_e^{H}$ (and obviously $n_e^{L}$) at these energies. Nevertheless, the $H$-resonance starts to be effective only for $E>70$ TeV, as $n^H_e$ falls below $10^{20}\, \text{cm}^{-3}$ and neutrinos produced at $n_e \approx 10^{21}\, \text{cm}^{-3}$ have the chance to cross the entire resonant layer. The upper limit on the effectiveness of the $H$-resonance is at $\sim 1$ PeV due to $\gamma$, in Eq.~\eqref{adiabaticity}, falling below $1$ for higher energies. The $L$-resonance, on the contrary, is effective even for energies down to $1$ TeV and up until $100$ PeV, thus covering the entire HEN-spectrum observed by IceCube~\cite{IceCube:2021uhz}. 

We can modify Eq.~\eqref{electron-density} either by changing its normalization, its power-law index, or both. For resonant conversion to happen, neutrinos must be produced in layers with densities of at least $n_e^L$.\footnote{Strictly speaking, even if neutrinos are not created in a region with densities $n_e^L$, as long as they cross a layer with such densities somewhere else during propagation, strong flavor conversion is still possible.} At the end of BLR ($ \sim 0.1$ pc), Eq.~\eqref{electron-density} is normalized to give $n_e=10^{9}$ $\text{cm}^{-3}$. However, even for normalization as small as $n_e=10$ $\text{cm}^{-3}$ (and power index $4$ to keep $n_e \geq n_e^L$ at the production region), resonant flavor conversion is still relevant. Therefore, the effect described here is  robust for a large range of profile parameters. On the other hand, profiles with indices $\lesssim 2$, although efficient for flavor conversion, could imply a BLR mass exceeding that of the central BH ($10^7  M_\odot$), potentially leading to gravitational instability.
\vspace{0.05in}

\section{Results} 
\begin{figure*}[htb!]
\includegraphics[width=0.99\textwidth]{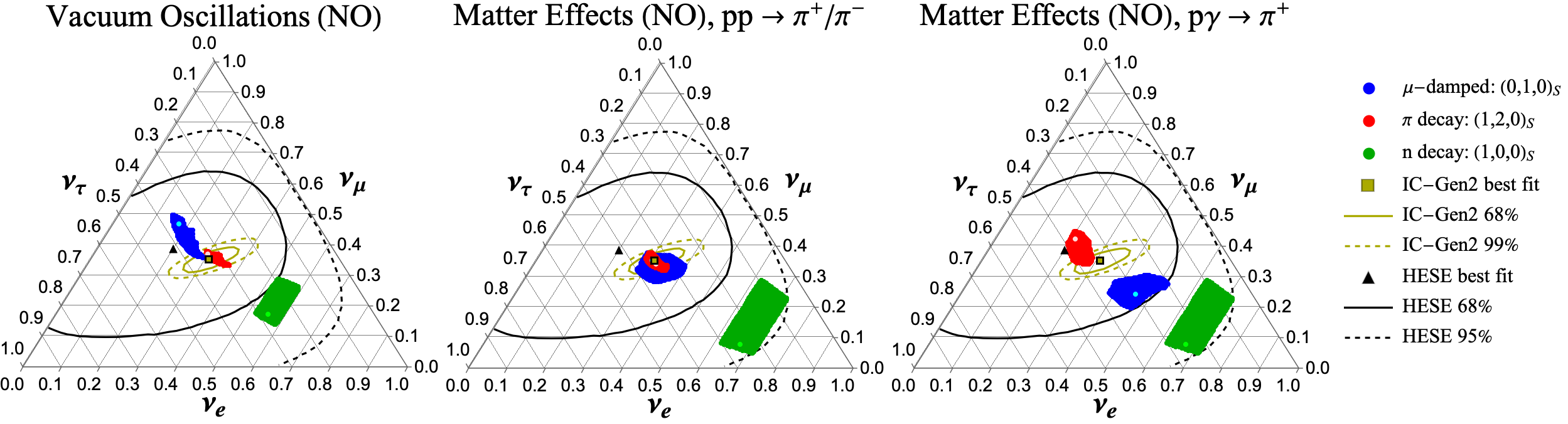}
  \caption{\textit{Left}: Allowed regions of flavor-ratios of astrophysical neutrinos on Earth after accounting for vacuum oscillations en route and the $3\sigma$ neutrino oscillation parameters (assuming normal ordering)  from a recent global fit~\cite{deSalas:2020pgw}.  \textit{Middle}: Same regions after the inclusion of ME at the source and assuming $pp$ production. \textit{Right}: Regions are further modified if instead we assume $p \gamma$ production in combination with ME. In each panel, we consider 3 scenarios: standard pion-decay (red), muon-damped (blue) and neutron decay (green). Also shown are the current IceCube constraints (black solid and dashed for 68\% and 95\% C.L. respectively)~\cite{IceCube:2020fpi}  and projections for IceCube-Gen2 (yellow)~\cite{IceCube-Gen2:2023rds}.}
\label{money_Fig}
\end{figure*}

In the absence of ME during propagation, the initial flavor-content is modified by averaged-out VO.\footnote{Due to the large distances and energy integration, the oscillatory terms of the VO probability cannot be resolved and are averaged out in the detector.} The probability that an initial flavor $\nu_\alpha$ will change to a flavor $\nu_\beta$ on its way to the Earth is given by~\cite{Pakvasa:2007dc, Chen:2014gxa} 
\begin{equation}
    P_{\alpha \beta}^{\mathrm{VO}}=\sum_{i=1}^3\left|U_{\alpha i}\right|^2\left|U_{\beta i}\right|^2 , 
\end{equation}
where $U_{\alpha i}$ are the PMNS matrix elements.
For an initial flavor composition $(f_{e}^{S},f_{\mu}^{ S},f_{\tau}^{ S})$ at the source, the composition at Earth is
\begin{equation}
    f_\beta^\oplus = \sum_{\alpha=e, \mu, \tau} P_{\alpha \beta}^{\mathrm{VO}} f_{\alpha}^{S} \, .
\end{equation}
Therefore, for VO without any ME, the flavor content for different production channels changes along propagation as (applicable for both $pp$ and $p\gamma$ sources) 
\begin{align}
\text{$\pi$-decay:}& \hspace{0.5 cm} (1/3,2/3,0)_S \rightarrow (0.3,0.37,0.33)_\oplus, \nonumber \\
\text{$\mu$-damped:}& \hspace{0.5 cm} (0,1,0)_S \rightarrow (0.17,0.47,0.36)_\oplus, \nonumber \\
\text{$n$-decay:}& \hspace{0.5 cm} (1,0,0)_S \rightarrow (0.55,0.17,0.28)_\oplus, 
\end{align}
using the best-fit values of the oscillation parameters~\cite{deSalas:2020pgw}, as shown by the colored circles on the left panel of Fig.~\ref{money_Fig}. Taking into account the $3\sigma$ interval of the parameters~\cite{deSalas:2020pgw} with flat priors, we get the colored regions in Fig.~\ref{money_Fig}. Results in the main text are shown for NO, while plots for IO, which are very similar to NO for VO, are given in Appendix~\ref{app:flavor}.  
Similar flavor-triangle analyses in the VO case have been done before~\cite{Lipari:2007su, Mena:2014sja, Palladino:2015zua, Bustamante:2015waa, Bustamante:2019sdb, Palladino:2019pid, Song:2020nfh}. 

Our main new result is that ME can substantially modify the allowed regions for flavor-ratios observed on Earth. First of all, different from VO, ME can distinguish neutrinos from antineutrinos and between production processes that are asymmetrical between them, such as $pp$ and $p \gamma$, see Fig.~\ref{money_Fig} middle and right panels. The $pp$ interactions produce roughly equal ratios of $\pi^+$ and $\pi^-$~\cite{Chen:2014gxa, Liu:2023lxz} and, therefore, $\nu_e$ and $\Bar{\nu}_e$ are created in equal amounts. On the other hand, $p \gamma$ suppresses the production $\pi^-$ relative to that of $\pi^+$, and hence, mostly $\nu_e$'s are created. Equal amounts of $\nu_\mu$ and $\Bar{\nu}_\mu$ are created in both processes. We have included the uncertainties in the ratio of $\pi^+/\pi^-$ production~\cite{Hummer:2010vx, Biehl:2016psj} to compute the regions shown in Fig.~\ref{money_Fig}. 

Here, we describe the pattern of flavor-conversion via $H$ and $L$-resonances in the central parts of AGNs using the energy level diagram in Fig.~\ref{Simplified_Fig} and assuming both layers to be perfectly adiabatic (see Appendix A and Ref.~\cite{Dighe:1999bi} for a detailed discussion of the energy level diagram). We start with $\nu_e$ ($\Bar{\nu}_e$), produced in dense matter as $\nu_3^m$ ($\nu_1^m$), which propagates to Earth as $\nu_3$ ($\nu_1$):
\begin{align}
    \text{For $\nu_e$:} \hspace{0.5 cm}(1,0,0)_S^\nu & \rightarrow\left(\left|U_{e 3}\right|^2,\left|U_{\mu 3}\right|^2,\left|U_{\tau 3}\right|^2\right)_{\oplus}^\nu \nonumber \\
& =(0.02,0.56,0.42)_{\oplus}^\nu. \label{nu_e} \\
\text{For $\Bar{\nu}_e$:} \hspace{0.5 cm}(1,0,0)_S^{\bar{\nu}} & \rightarrow\left(\left|U_{e 1}\right|^2,\left|U_{\mu 1}\right|^2,\left|U_{\tau 1}\right|^2\right)_{\oplus}^{\bar{\nu}} \nonumber \\
& =(0.67,0.08,0.25)_{\oplus}^{\bar{\nu}}.  \label{nu_e_bar}
\end{align}
The superscripts $\nu$ and $\Bar{\nu}$ indicate that a given flavor is populated only by neutrinos or antineutrinos. Similarly, $\nu_\mu$ ($\Bar{\nu}_\mu$) propagates to Earth as incoherent mixture of $\nu_1$ and $\nu_2$ ($\nu_2$ and $\nu_3$):
\begin{align}
&   \text{For $\nu_\mu$:} \hspace{0.5 cm}(0,1,0)_S^\nu \rightarrow \nonumber  \\
&\frac{1}{2}\left(\left|U_{e 1}\right|^2+\left|U_{e 2}\right|^2,\left|U_{\mu 1}\right|^2+\left|U_{\mu 2}\right|^2,\left|U_{\tau 1}\right|^2+\left|U_{\tau 2}\right|^2 \right)_{\oplus}^\nu 
\nonumber  \\
 & \qquad \qquad =(0.5,0.2,0.3)_\oplus^\nu . \label{nu_mu} \\
 &\text{For $\Bar{\nu}_\mu$:} \hspace{0.5 cm}(0,1,0)_S^{\Bar{\nu}} 
 \rightarrow \nonumber  \\
& \frac{1}{2}\left(\left|U_{e 2}\right|^2+\left|U_{e 3}\right|^2,\left|U_{\mu 2}\right|^2+\left|U_{\mu 3}\right|^2,\left|U_{\tau 2}\right|^2+\left|U_{\tau 3}\right|^2 \right)_{\oplus}^{\Bar{\nu}}
\nonumber  \\
 & \qquad \qquad =(0.17,0.46,0.37)_{\oplus}^{\Bar{\nu}}. \label{nu_mu_bar}
\end{align}
With Eqs.~\eqref{nu_e}-\eqref{nu_mu_bar} we can estimate the results for all relevant production processes of HENs in AGNs, taking ME into account. For $pp$ process, total $\pi^+/\pi^-$ decay leads to the proportion $f_{\nu_e}^S=f_{\bar{\nu}_e}^S=\frac{1}{6},f_{\nu_\mu}^S=f_{\bar{\nu}_\mu}^S=\frac{1}{3},f_{\nu_\tau}^S=f_{\bar{\nu}_\tau}^S=0,$
while for $\mu$-damped case, we have $f_{\nu_\mu}^S=f_{\bar{\nu}_\mu}^S=\frac{1}{2},f_{\nu_e}^S=f_{\nu_\tau}^S=f_{\bar{\nu}_e}^S=f_{\bar{\nu}_\tau}^S=0$. Taking the corresponding weighted averages with Eqs.~\eqref{nu_e}-\eqref{nu_mu_bar}, we get  
\begin{align}
    \text{$\pi$-decay ($pp$):} & \hspace{0.5 cm} (1/3,2/3,0)_S \rightarrow (0.34,0.33,0.33)_\oplus. \nonumber \\
    \text{$\mu$-damped ($pp$):} & \hspace{0.5 cm} (0,1,0)_S \rightarrow (0.34,0.33,0.33)_\oplus.
\end{align}
Contrary to VO case, for ME and $pp$ production, the $\pi$-decay and $\mu$-damped processes cannot be disentangled, see Fig.~\ref{money_Fig} (middle panel).

For $p \gamma$ case, only $\pi^+$ is produced from $\Delta^+$ resonance and decays in the proportion $f_{\nu_e}^S=f_{\nu_\mu}^S=f_{\bar\nu_\mu}^S=\frac{1}{3}$, $f_{\nu_\tau}^S=f_{\bar\nu_e}^S=f_{\bar\nu_\tau}^S=0$, while $\mu$-damped case creates only $\nu_\mu$. Following the same weighted average procedure, we get
\begin{align}
    \text{$\pi$-decay ($p \gamma$):}&  \hspace{0.5 cm} (1/3,2/3,0)_S \rightarrow (0.23,0.4,0.37)_\oplus. \nonumber \\
    \text{$\mu$-damped ($p \gamma$):}& \hspace{0.5 cm} (0,1,0)_S \rightarrow (0.5,0.2,0.3)_\oplus.
\end{align}
In this case, both $\pi$-decay and $\mu$-damped are completely displaced from their VO expected regions in the flavor-triangle, see Fig.~\ref{money_Fig} right panel. In particular, $\mu$-damped and ME can be confused with $n$-decay and VO, while  $\pi$-decay and ME can be confused with $\mu$-damped and VO.

For both processes,
\begin{align}
    \text{$n$-decay ($pp$ \& $p \gamma$):}& ~ (1,0,0)_S \rightarrow (0.67,0.08,0.25)_\oplus.
\end{align}
The allowed region corresponding to $n$-decay is also displaced to the lower right vertex on the flavor-triangle when compared to the pure VO case. The most impressive change in the $n$-decay allowed region comes in IO, see Appendix~\ref{app:flavor}. We summarize the best-fit NO results in Table~\ref{table-NO}. 

\begin{table}[t!]
\centering
\begin{tabular}{|c|c|}
\hline \hline \multicolumn{2}{|c|}{\textbf{Vacuum Oscillations (NO)}} \\
\hline$\pi$-decay & $\hspace{0.5 cm} (1/3,2/3,0)_S \rightarrow (0.30,0.37,0.33)_\oplus \hspace{0.5 cm}$ \\
\hline$\mu$-damped & $\hspace{0.5 cm} (0,1,0)_S \rightarrow (0.17,0.47,0.36)_\oplus \hspace{0.5 cm}$ \\
\hline$n$-decay & $\hspace{0.5 cm} (1,0,0)_S \rightarrow (0.55,0.17,0.28)_\oplus \hspace{0.5 cm}$\\
\hline
\end{tabular}

\begin{tabular}{|c|c|}
\hline \multicolumn{2}{|c|}{\textbf{Matter Effect (NO)}, $p p$ production} \\
\hline$\pi$-decay & $\hspace{0.5 cm} (1/3,2/3,0)_S \rightarrow (0.34,0.33,0.33)_\oplus \hspace{0.5 cm}$ \\
\hline$\mu$-damped & $\hspace{0.5 cm} (0,1,0)_S \rightarrow (0.34,0.33,0.33)_\oplus \hspace{0.5 cm}$ \\
\hline$n$-decay & $\hspace{0.5 cm} (1,0,0)_S \rightarrow (0.67,0.08,0.25)_\oplus \hspace{0.5 cm}$\\
\hline
\end{tabular}

\begin{tabular}{|c|c|}
\hline \multicolumn{2}{|c|}{\textbf{Matter Effect (NO)}, $p \gamma$ production} \\
\hline$\pi$-decay & $\hspace{0.5 cm} (1/3,2/3,0)_S \rightarrow (0.23,0.40,0.37)_\oplus \hspace{0.5 cm}$ \\
\hline$\mu$-damped & $\hspace{0.5 cm} (0,1,0)_S \rightarrow (0.50,0.20,0.30)_\oplus \hspace{0.5 cm}$ \\
\hline$n$-decay & $\hspace{0.5 cm} (1,0,0)_S \rightarrow (0.67,0.08,0.25)_\oplus \hspace{0.5 cm}$\\
\hline
\hline
\end{tabular}
\caption{Summary of our flavor-conversion results, assuming normal mass ordering.}
\label{table-NO}
\end{table}

Fig.~\ref{money_Fig} also includes the $68\%$ and $95\%$ C.L. bounds on the flavor composition of the diffuse HEN-flux from the IceCube analysis~\cite{IceCube:2020fpi} using the High-Energy Starting Events (HESE) sample~\cite{IceCube:2020wum}, which is an all-sky and all-flavor search with neutrinos above $60$ TeV, collected during 7.5 years of IceCube lifetime. Observe that while the $95 \%$ C.L. region is compatible with all possible flavor compositions, the current best fit is closer to $\mu$-damped region for VO. After including ME and assuming dominant AGN contribution to the HEN flux~\cite{IceCube:2021pgw}, the best fit can be explained by a pure $\pi$-decay after $p \gamma$ production, see Fig.~\ref{money_Fig}. In this last scenario, all other production processes are partially outside of the $68 \%$ C.L., although they can be relevant in combination with $\pi$-decay. Nevertheless, more statistics is necessary to further constrain the flavor-ratios, see, for example, the preliminary analysis in Ref.~\cite{IceCube:2023fgt}.
Future detectors will also improve our understanding of the flavor composition in the next decades~\cite{Song:2020nfh}. In Fig.~\ref{money_Fig}, we include projections from IceCube-Gen2 after $10$ years of operations (yellow contours) assuming $\pi$-decay and VO to be the true hypothesis~\cite{IceCube-Gen2:2023rds}.

Neutrino flavor-conversion in AGNs can also be directly studied by using the point-source flux from identified sources, as opposed to the all-sky diffused flux. The increasing evidence for neutrinos pointing to steady-state AGN sources can allow for individual studies of flavor composition in different energy regimes with future data. 
\vspace{0.05in}

\section{Discussion and Conclusions}
We have examined the resonant flavor-conversion of HENs within the {\it standard 3-neutrino oscillation paradigm}. The AGN model used in Eq.~\eqref{electron-density} has a column density $N_{\rm H}\equiv \int n_e dr \sim 10^{33}\, \text{cm}^{-2}$, compatible with the minimum necessary width of the medium for strong flavor-conversion~\cite{Lunardini:2000swa}. Flavor conversion via the $L$-resonance alone can be achieved at smaller $N_{\rm H}\sim 10^{30} \, \text{cm}^{-2}$, which is still orders of magnitude higher than that ($\sim 10^{25}\, \text{cm}^{-2}$) inferred for NGC 1068 from $X$-ray studies~\cite{Bauer:2014rla}. Such heavily Compton-thick AGNs, with $N_{\rm H}\geq \sigma_T^{-1}\simeq 1.5\times 10^{24}\, \text{cm}^{-2}$ (inverse of the Thomson cross-section, which corresponds to unity optical depth for Compton scattering), could be numerous~\cite{Carroll:2023mos}, but detecting them is challenging by conventional astrophysical methods due to $X$-ray absorption~\cite{Hickox:2018xjf}. HENs, unaffected by obscuration, offer an exciting multi-messenger avenue to study these AGNs, complementary to searches based on electromagnetic emission and star formation. Our proposed mechanism, measuring a shift in the flavor-ratios influenced by ME, provides a unique probe of the heavily Compton-thick AGNs as the sources of the HESE neutrinos, while also offering a probe for transient emissions caused by binary merger objects embedded in ADs \cite{Yuan:2021bxx}. Understanding the distribution of Compton-thick AGNs is also crucial for modeling their impact on the cosmic $X$-ray background and gaining insights into the correlation between black hole growth and galaxy evolution~\cite{Hickox:2018xjf}.


\section{Acknowledgments} We thank Manel Errando and Alexei Smirnov for useful discussions. The work of BD was partly supported by the U.S. Department of Energy under grant No. DE-SC 0017987. The work of YP
was supported in part by the S\~{a}o Paulo Research Foundation (FAPESP) Grant No. 2023/10734-3. We thank the Fermilab Theoretical Physics Department, where part of this work was done, for their warm hospitality. BD and SJ also wish to acknowledge the Center for Theoretical Underground Physics and Related Areas (CETUP*) and the Institute for Underground Science at SURF for hospitality and for providing a stimulating environment. 

\appendix
\section{Energy levels} 
\label{app:energy}
Here, we discuss the energy levels of the Hamiltonian in Eq.~3, which we show in Fig.~1 for normal ordering (NO). For illustrative purposes, these energy levels are plotted for fictitious values of $\Delta m^2_{21}$ and $\Delta m^2_{31}$ ($7$ $\text{eV}^2$ and $15$ $\text{eV}^2$, respectively)\footnote{This is a standard practice in the literature~\cite{Dighe:1999bi, Akhmedov:2003fu}.} while the mixing angles are chosen close to their best-fit values ($\theta_{12}=34^\circ$, $\theta_{13}=8^\circ$, $\theta_{23}=45^\circ$).

Firstly, recall from the previous discussion that we can represent antineutrinos as neutrinos with negative $V_e$. At very high densities, or large $|V_e|$, $|\mathcal{H}_{ee}|$ is much larger than the mixing terms in $\mathcal{H}$ and $\nu_e$ ($\Bar{\nu}_e$) becomes the heaviest (lightest) matter eigenstate in the medium. Meanwhile, the other eigenstates are roughly an equal mixture ($\theta_{23} \sim 45^o$) of $\nu_\mu$ and $\nu_\tau$; we denote them as $\nu'_\mu$ and $\nu'_\tau$. Thus, 
\begin{align}
   &  \nu_3^m=\nu_e \, , \quad \nu_2^m=\nu'_\tau \, , \quad \nu_1^m=\nu'_\mu \, , \\
   & \bar{\nu}_1^m = \bar{\nu}_e \, , \quad \bar{\nu}_2^m= \bar{\nu}'_\mu \, , \quad \bar{\nu}_3^m=\bar{\nu}'_\tau \, , 
\end{align}
where the superscript $m$ stands for `matter'.

As $\mathcal{H}_{ee}$ is linear in $V_{ee}$, for $\nu_e$ ($\Bar{\nu}_e$) produced at high densities $n_e\gg n_H^{\rm res},n_L^{\rm res}$, their energy levels can cross the levels of $\nu'_\mu$ and $\nu'_\tau$ ($\bar\nu'_\mu$ and $\bar\nu'_\tau$) as they propagate towards vacuum densities $n_e\ll n_H^{\rm res},n_L^{\rm res}$. At these crossings, the $H$ and $L$-resonances occur~\cite{Dighe:1999bi}. For NO, both resonances happen for neutrinos. For IO, the $H$-resonance moves to the antineutrino channel. While the levels of the flavor states cross, the levels of matter eigenstates ($\nu_3^m$, $\nu_2^m$, and $\nu_1^m$) do not. Therefore, the eigenstates in matter exchange flavor content at resonances and lead the system to flavor conversion.


\section{{Clumpy BLR}}
\label{app:BLR}

We clarify the conditions under which the smooth-medium approximation used in the main text is valid even when the BLR is composed of discrete clouds. We first review the physics of resonance in a smooth medium (Sec.~\ref{app:B1}), then describe how a clumpy BLR  modifies this picture Sec.~\ref{app:B2}). In Sec.~\ref{app:B3}, we explain the two physically distinct scenarios in which the smooth approximation remains valid. In Sec.~\ref{app:B4}, we discuss the physical conditions under which AGNs can realize either of these two distinct scenarios and how this is corroborated by observations.

\subsection{Resonance shell and adiabaticity in a smooth medium}
\label{app:B1}

Neutrino evolution in matter is governed by the effective mixing angle 
\begin{equation}
\tan 2\theta_m = \frac{\sin 2\theta}{\cos 2\theta - A(r)} \, ,
\end{equation}
where $\theta$ is the vacuum mixing angle in the two-flavor approximation, and
\begin{equation}
A(r) = \frac{2E V_e(r)}{\Delta m^2},
\end{equation}
with $V_e(r) = \sqrt{2} G_F n_e(r)$ being the matter potential. At the resonance radius $r_{\rm res}$, defined by the condition $A(r_{\rm res}) = \cos 2\theta$ [see Eq.~\eqref{resonance-conditions}], the matter mixing angle becomes maximal, $\theta_m = \pi/4$. Conversion is efficient not only at this point but also within a finite resonance shell, defined by the condition $|A - \cos 2\theta| \lesssim \sin 2\theta$. Expanding $A(r)$ around $r_{\rm res}$, the thickness of this shell is
\begin{equation}
\Delta r_{\rm res}\simeq \frac{\sin 2\theta}{|dA/dr|_{\rm res}}
=\tan 2\theta\;\, \left|\frac{n_e}{dn_e/dr}\right|.
\end{equation}

In the next subsection, we extend this rationale to the case of a clumpy medium, describing it in terms of its average density.

\subsection{Clumpy BLR: Local description with radial dependence}
\label{app:B2}

In a clumpy BLR, the relevant cloud parameters may vary with distance from the central supermassive BH. We denote by $n_e^{ c}(r)$ the electron density inside the clouds, by $n_e^{\rm ic}(r)$ the density of the inter-cloud medium (assumed negligible), and by $f(r) \in [0,1]$ the volume filling factor, which corresponds to the fraction of space along the line of sight that is occupied by clouds. Throughout this discussion, we assume that all these quantities vary radially.
The mean density at radius $r$ is then
\begin{equation}
\langle n_e(r) \rangle=f(r)\,n_e^{c}(r)+\big[1-f(r)\big]\,n_e^{\rm ic}(r).
\end{equation}
For a neutrino of energy $E$, the resonance condition is satisfied at the radius $r_{\rm res}$ such that
\begin{equation}
\langle n_e(r_{\rm res}) \rangle=n_e^{\rm res}=\frac{\Delta m^2}{2\sqrt{2}\,G_F\,E}\cos 2\theta.
\end{equation}
The associated resonance shell, within which mixing is large, has thickness
\begin{equation}
\Delta r_{\rm res}\simeq \tan 2\theta \,  \frac{\langle n_e(r_{\rm res})\rangle}{\big|d\langle n_e(r)\rangle/dr\big|_{r_{\rm res}}}.
\end{equation}
Even though this average-density description is intuitive, in practice, it is valid in certain limits of the clumpy BLR, as discussed in the next subsection. 

\subsection{Local validity conditions} \label{app:B3}

The smooth-profile BLR results derived in the main text remain valid if either of the following criteria is satisfied locally at the resonance shell.

\begin{itemize}

\item \textbf{Condition A: Adiabatic edges.}  
Along neutrino propagation, each cloud boundary produces a density change $|\Delta n_e|$ across an edge of thickness $\delta \ell_c$. Approximating the gradient as $|dn_e/dr| \simeq |\Delta n_e| / \delta \ell_c$ in Eq.~\eqref{adiabaticity}, the adiabaticity requirement $\gamma > 1$ implies
\begin{equation}
\delta \ell_c(r_{\rm res}) >
\frac{2E\,\cos 2\theta}{\Delta m^2\sin^2 2\theta}\;
\frac{|\Delta n_e(r_{\rm res})|}{n_e^{\rm res}}.
\end{equation}
If this condition holds, the neutrino tracks its eigenstate smoothly across each edge, even if its trajectory contains only a few clouds.

\item \textbf{Condition B: Many clouds and small fluctuations in density.}  
To a good approximation, the expected number of clouds within the resonance shell is
\begin{equation} \label{requirement}
N_{c}^{\rm res} \simeq
\frac{f(r_{\rm res})\,\Delta r_{\rm res}}{\ell_c(r_{\rm res})},
\end{equation}
where $\ell_c(r_{\rm res})$ denotes the typical cloud radius at the resonance shell.
The average-density approximation of Sec.~\ref{app:B2} can be applied if the neutrino trajectory satisfies $N_{c}^{\rm res} \gg 1$, together with the additional requirements
\begin{equation} \label{add-requirement}
\frac{\sigma_{n_e}}{\langle n_e\rangle}\;\ll\;\tan 2\theta,
\qquad
\ell_c, \ell_{\rm ic} \ll L_{\rm osc}^{\rm res},
\end{equation}
where $\sigma_{n_e}$ is the rms density fluctuation, $\ell_{\rm ic}$ is the inter-cloud separation, and $L_{\rm osc}^{\rm res}$ is the oscillation length at resonance. All these quantities should be evaluated at the resonance shell.
These conditions ensure that fluctuations remain within the resonance width and occur on scales too short to disrupt coherent MSW conversion.

Ideally, Eqs.~\eqref{requirement} and \eqref{add-requirement} should hold throughout the entire neutrino propagation, not only at the resonance shell. In a minimal scenario, however, neutrinos are produced at a density just above the $L$-resonance and then only have to traverse the resonance shell before reaching vacuum. Therefore, the description in terms of the resonance layer given above provides the minimal conditions for Condition B to be realized.
\end{itemize}

If neither condition A nor B is satisfied — i.e., edges are sharp and the resonance shell contains only a few segments — the evolution becomes non-adiabatic, with one or more Landau–Zener transitions of probability $P_c \simeq e^{-\pi\gamma/2}$. The final flavor outcome then depends on the sequence of jumps and the phases between them. In this regime, MSW conversion is reduced, and the flavor composition can lie anywhere between the adiabatic-MSW and vacuum-averaged limits, depending on the local structure of the density profile.

\subsection{Validity conditions in real BLRs}
\label{app:B4}

In this subsection, we assume the simple scenario in which neutrinos are produced just above the $L$-resonance and therefore only have to cross one resonant layer before reaching vacuum. For the $L$-resonant layer at $E = 100~\text{TeV}$, the quantities of interest are $L^{\rm res}_{\rm osc}(L) \simeq 10^{-4}~\text{pc}$,
$\Delta r_{\rm res}(L) \simeq  10^{-3}~\text{pc}$,
and $\delta \ell_c^{\min}(L) \simeq 5.6\times 10^{-6}~\text{pc}$.

We begin by considering observations of NGC 4395, the lowest-luminosity known AGN, whose BLR is extremely compact ($\ell_{\rm tot} \sim 10^{-4}$~pc). These observations are consistent with a BLR composed of a very large number of small clouds, a configuration that could realize Condition~B.
The absence of large-amplitude fluctuations in the emission-line profiles of NGC 4395 can be translated into a lower bound on the total number of BLR clouds, $N_{c}^{\rm tot} \gtrsim 10^4$, and a corresponding upper bound on their typical size, $\ell_c \lesssim 3 \times 10^{-7}$~pc~\cite{Laor:2005gg}. Here $N_{c}^{\rm tot}$ refers to the total number of clouds filling the entire BLR volume, in contrast to $N_{c}^{\rm res}$ [Eq.~\eqref{requirement}], which denotes only the number of clouds intersecting the resonant layer along the neutrino trajectory.
  
So let us assume BLR cloud sizes of order $\ell_c \lesssim 10^{-6}$~pc in the resonant layer. In this case,
  \[
  N_{c}^L \;\simeq\; \frac{f\,\Delta r_{\rm res}(L)}{\ell_c(r_{\rm res})}
  \gtrsim f\times 10^3.
  \]
Thus, Condition~B is satisfied, i.e., $N_{c} \gg 1$, provided $f \gg 10^{-3}$. Such a filling factor is consistent with the values reported in Ref.~\cite{Laor:2005gg}. To estimate it, we compute the fraction of the total BLR volume that is occupied by clouds:
\begin{align}
      f \approx \frac{N_{c}^{\rm tot} \times V_{c}}{V_{\rm BLR}} \simeq N_{c}^{\rm tot} \times \left(\frac{\ell_c}{\ell_{\rm BLR}} \right)^3.
\end{align}
Here we assume, for simplicity, that the clouds are spherical, of the same size, and uniformly distributed. For $\ell_c \approx 10^{-6}~\text{pc}$ and $N_{c}^{\rm tot} > 10^4$, this gives $f \gtrsim 10^{-2}$. Therefore, Condition~B, namely $N_{c}^L \gg 1$, lies well within the realm of possibilities for real BLRs.

BLRs with typical sizes up to $0.1$~pc, as assumed in the derivations of our main results, may also host numerous small clouds that could help to sustain a large volume filling factor. However, even for very small filling factors, where Condition~B cannot be satisfied, Condition~A may still hold for large and smooth clouds whose edges extend over $\delta \ell_c > \delta \ell_c^{\min}(L) \simeq 5.6 \times 10^{-6}$~pc. This requirement is achievable, since the necessary edge thickness is much smaller than the typical cloud sizes quoted in the literature, $\sim 10^{-4}$~pc~\cite{Armijos-Abendano:2022upj}.

Therefore, our results remain valid even in a clumpy BLR scenario, provided certain values of the BLR cloud parameters are realized -- a scenario possible for real AGNs as illustrated above. Correspondingly, we expect that at least a fraction of AGNs could emit neutrinos that are consistently affected by sizable matter effects discussed in the main text, irrespective of whether the BLRs are discrete or continuous.

\section{More Flavor Triangles}
\label{app:flavor}
\begin{figure*}[htb!]
\includegraphics[width=0.99\textwidth]{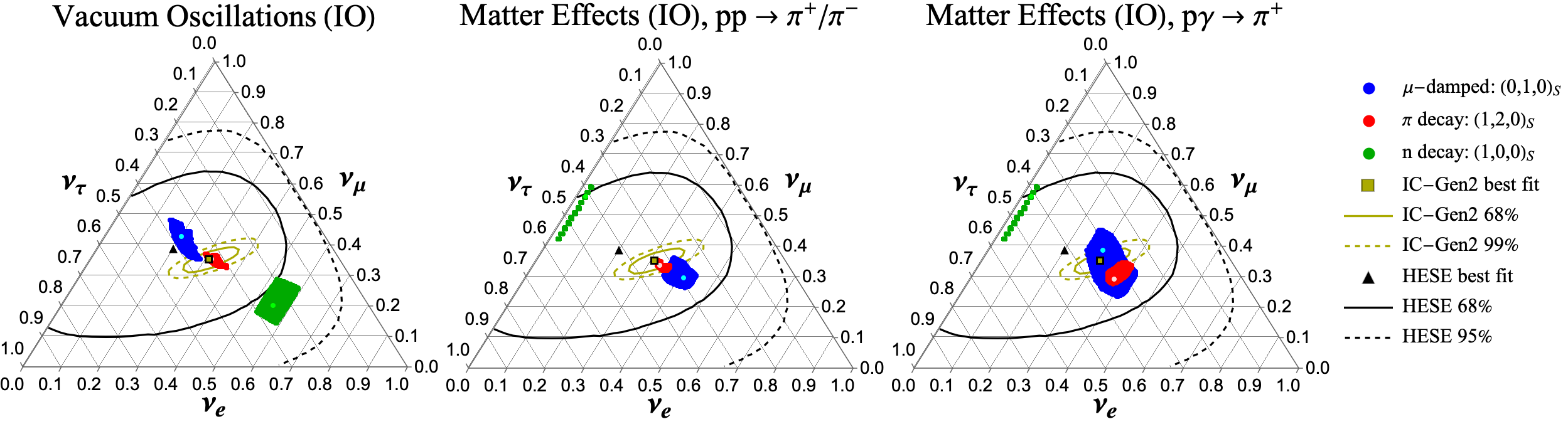}
  \caption{Same as Fig.~2 but for inverted ordering (IO).}
\label{money_Fig_IO}
\end{figure*}

\begin{figure*}[htb!]
\includegraphics[width=0.99\textwidth]{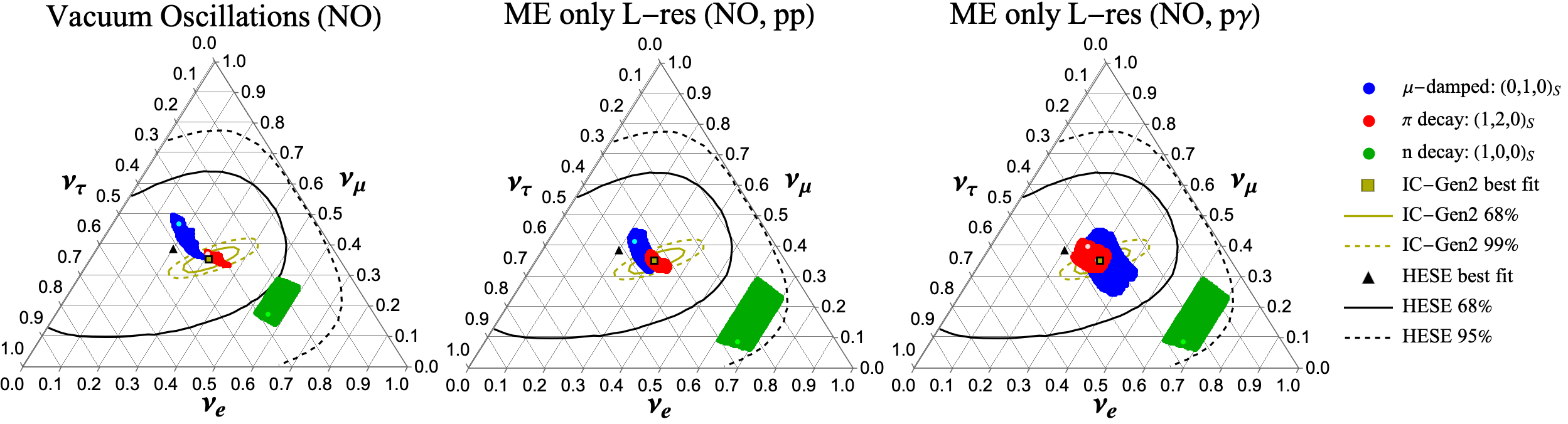}
  \caption{Same as Fig.~2 (NO) but assuming that only the $L$-resonance is effective.}
\label{money_Fig_NO_L}
\end{figure*}

\begin{figure*}[htb!]
\includegraphics[width=0.99\textwidth]{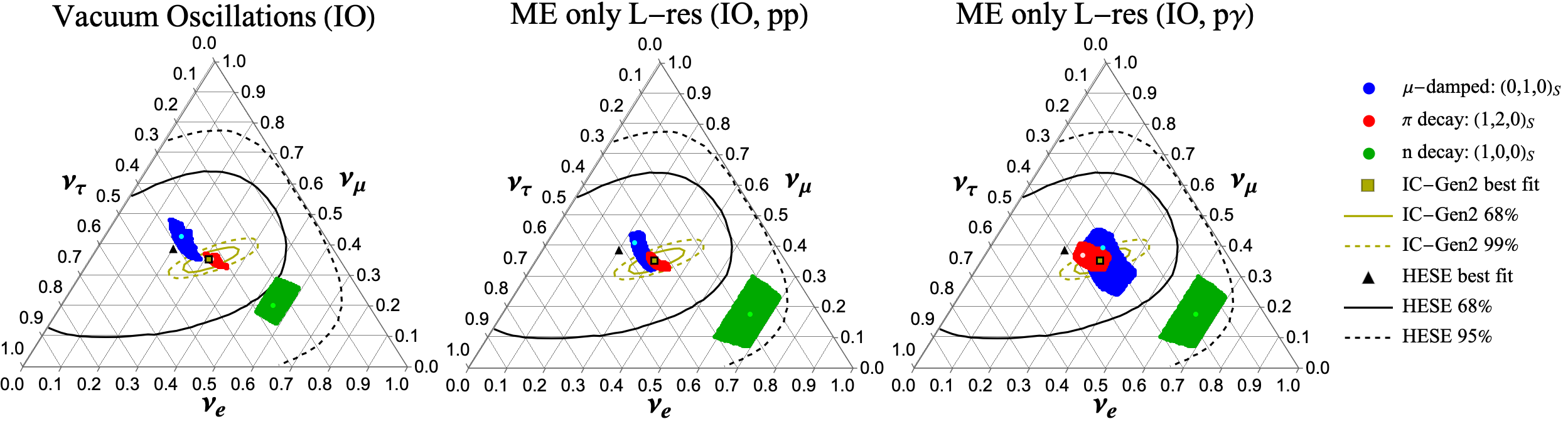}
  \caption{Same as Fig.~3 (IO) but assuming that only the $L$-resonance is effective.}
\label{money_Fig_IO_L}
\end{figure*}

The flavor triangle analysis, including the ME for the IO scenario, is shown in Fig.~\ref{money_Fig_IO}, and summarized in Table~\ref{table-IO}. Finally, Figs.~\ref{money_Fig_NO_L} and \ref{money_Fig_IO_L} show the results for NO and IO, respectively, for the case where only the $L$-resonance is effective. 

\begin{table}[htb!]
\centering
\begin{tabular}{|c|c|}
\hline \hline \multicolumn{2}{|c|}{\textbf{Vacuum Oscillations (IO)}} \\
\hline$\pi$-decay & $\hspace{0.5 cm} (1/3,2/3,0)_S \rightarrow (0.32,0.35,0.33)_\oplus \hspace{0.5 cm}$ \\
\hline$\mu$-damped & $\hspace{0.5 cm} (0,1,0)_S \rightarrow (0.20,0.42,0.38)_\oplus \hspace{0.5 cm}$ \\
\hline$n$-decay & $\hspace{0.5 cm} (1,0,0)_S \rightarrow (0.55,0.20,0.25)_\oplus \hspace{0.5 cm}$\\
\hline
\end{tabular}

\begin{tabular}{|c|c|}
\hline \multicolumn{2}{|c|}{\textbf{Matter Effect (IO)}, $p p$ production} \\
\hline$\pi$-decay & $\hspace{0.5 cm} (1/3,2/3,0)_S \rightarrow (0.33,0.33,0.34)_\oplus \hspace{0.5 cm}$ \\
\hline$\mu$-damped & $\hspace{0.5 cm} (0,1,0)_S \rightarrow (0.42,0.28,0.30)_\oplus \hspace{0.5 cm}$ \\
\hline$n$-decay & $\hspace{0.5 cm} (1,0,0)_S \rightarrow (0.02,0.56,0.42)_\oplus \hspace{0.5 cm}$\\
\hline
\end{tabular}

\begin{tabular}{|c|c|}
\hline \multicolumn{2}{|c|}{\textbf{Matter Effect (IO)}, $p \gamma$ production} \\
\hline$\pi$-decay & $\hspace{0.5 cm} (1/3,2/3,0)_S \rightarrow (0.38,0.28,0.34)_\oplus \hspace{0.5 cm}$ \\
\hline$\mu$-damped & $\hspace{0.5 cm} (0,1,0)_S \rightarrow (0.34,0.36,0.30)_\oplus \hspace{0.5 cm}$ \\
\hline$n$-decay & $\hspace{0.5 cm} (1,0,0)_S \rightarrow (0.02,0.56,0.42)_\oplus \hspace{0.5 cm}$\\
\hline
\hline
\end{tabular}
\caption{Summary of our flavor conversion results, assuming inverted mass ordering.}
\label{table-IO}
\end{table}

\bibliographystyle{utcaps_mod}
\bibliography{reference}

\end{document}